\documentclass[journal]{IEEEtran}

\usepackage[numbers,sort&compress]{natbib}
\usepackage{url}
\usepackage{bm}
\usepackage{amsfonts}
\usepackage{amsmath}
\usepackage{amssymb}
\usepackage{amsthm}
\usepackage{color}
\usepackage{enumerate}

\usepackage{array,color}
\usepackage{tabularx}
\usepackage{multirow}
\usepackage{booktabs}
\usepackage{makecell}  

\usepackage{graphicx}  
\usepackage{subfig}
\usepackage{epstopdf}
\usepackage{graphics}
\usepackage{epsfig}
\usepackage{psfrag}
\usepackage{overpic}

 \usepackage{stfloats}
 \usepackage{float}   

 \usepackage{indentfirst} 
\usepackage{gensymb}

\begin{document}

\newcommand {\Data} [1]{\mbox{${#1}$}}  

\newcommand {\DataN} [2]{\Data{\Power{{#1}}{{{#2}}}}}  
\newcommand {\DataIJ} [3]{\Data{\Power{#1}{{{#2}\!\times{}\!{#3}}}}}  

\newcommand {\DatassI} [2]{\!\Data{\Index{#1}{\!\Data 1},\!\Index{#1}{\!\Data 2},\!\cdots,\!\Index{#1}{\!{#2}}}}  
\newcommand {\DatasI} [2]{\Data{\Index{#1}{\Data 1},\Index{#1}{\Data 2},\cdots,\Index{#1}{#2},\cdots}}   
\newcommand {\DatasII} [3]{\Data{\Index{#1}{{\Index{#2}{\Data 1}}},\Index{#1}{{\Index{#2}{\Data 2}}},\cdots,\Index{#1}{{\Index{#2}{#3}}},\cdots}}  

\newcommand {\DatasNTt}[3]{\Data{\Index{#1}{{#2}{\Data 1}},\Index{#1}{{#2}{\Data 2}},\cdots,\Index{#1}{{#2}{#3}}} } 
\newcommand {\DatasNTn}[3]{\Data{\Index{#1}{{\Data 1}{#3}},\Index{#1}{{\Data 2}{#3}},\cdots,\Index{#1}{{#2}{#3}}} } 

\newcommand {\Vector} [1]{\Data {\mathbf {#1}}}
\newcommand {\Rdata} [1]{\Data {\hat {#1}}}
\newcommand {\Tdata} [1]{\Data {\tilde {#1}}} 
\newcommand {\Udata} [1]{\Data {\overline {#1}}} 
\newcommand {\Fdata} [1]{\Data {\mathbb {#1}}} 
\newcommand {\Prod} [2]{\Data {\prod_{\SI {#1}}^{\SI {#2}}}}  
\newcommand {\Sum} [2]{\Data {\sum_{\SI {#1}}^{\SI {#2}}}}   
\newcommand {\Belong} [2]{\Data{ {#1} \in{}{#2}}}  

\newcommand {\Abs} [1]{\Data{ \lvert {#1} \rvert}}  
\newcommand {\Mul} [2]{\Data{ {#1} \times {#2}}}  
\newcommand {\Muls} [2]{\Data{ {#1} \! \times \!{#2}}}  
\newcommand {\Mulsd} [2]{\Data{ {#1} \! \cdot \!{#2}}}  
\newcommand {\Div} [2]{\Data{ \frac{#1}{#2}}}  
\newcommand {\Trend} [2]{\Data{ {#1}\rightarrow{#2}}}  
\newcommand {\Sqrt} [1]{\Data {\sqrt {#1}}} 
\newcommand {\Sqrnt} [2]{\Data {\sqrt[2]{#1}}} 

\newcommand {\Power} [2]{\Data{ {#1}^{\TI {#2}}}}  
\newcommand {\Index} [2]{\Data{ {#1}_{\TI {#2}}}}  

\newcommand {\Equ} [2]{\Data{ {#1} = {#2}}}  
\newcommand {\Equs} [2]{\Data{ {#1}\! =\! {#2}}}  
\newcommand {\Equss} [3]{\Equs {#1}{\Equs {#2}{#3}}}  

\newcommand {\Equu} [2]{\Data{ {#1} \equiv {#2}}}  

\newcommand {\LE}[0] {\leqslant}
\newcommand {\GE}[0] {\geqslant}
\newcommand {\NE}[0] {\neq}
\newcommand {\INF}[0] {\infty}
\newcommand {\MIN}[0] {\min}
\newcommand {\MAX}[0] {\max}

\newcommand {\Funcfx} [2]{\Data{ {#1}({#2})}}  
\newcommand {\Funcfzx} [3]{\Data{ {\Index {#1}{#2}}({#3})}}  
\newcommand {\Funcfnzx} [4]{\Data{ {\Index {\Power{#1}{#2}}{#3}}({#4})}}  
\newcommand {\SI}[1] {\small{#1}}
\newcommand {\TI}[1] {\tiny {#1}}
\newcommand {\Text}[1] {\text {#1}}

\newcommand {\VtS}[0]{\Index {t}{\Text {s}}}
\newcommand {\Vti}[0]{\Index {t}{i}}
\newcommand {\Vt}[0]{\Data {t}}
\newcommand {\VLES}[1]{\Index {\tau} {\SI{\Index {}{ \Text{#1}}}}}
\newcommand {\VLESmin}[1]{\Index {\tau} {\SI{\Index {}{ \Text{min\_\Text{#1}}}}}}
\newcommand {\VAT}[0]{\Index {\Vector A}{\Text{Time}}}
\newcommand {\VPbus}[1]{\Index {P}{\Text{Node-}{#1}}}
\newcommand {\VPbusmax}[1]{\Index {P}{\Text{max\_Node-}{#1}}}

\newcommand {\EtS}[2]{\Equs {\Index {t}{\Text {s}}}{#1} {#2}}
\newcommand {\Eti}[2]{\Equs {\Index {t}{i}}{#1} {#2}}
\newcommand {\Et}[2]{\Equs {t}{#1} {#2}}
\newcommand {\EMSR}[2]{\Equs {\Index {\tau} {\SI{\Index {}{ \Text{MSR}}}}}{#1} {#2}}
\newcommand {\EMSRmin}[2]{\Equs {\Index {\tau} {\SI{\Index {}{ \Text{MSR}}}}}{#1} {#2}}

\newcommand {\EAT}[2]{\Equs {\Index {\Vector A}{\Text{Time}}}{#1} {#2}}
\newcommand {\EPbus}[3]{\Equs {\Index {P}{\Text{Node-}{#1}}}{#2} \Text{ #3}}
\newcommand {\EPbusmax}[3]{\Equs {\Index {P}{\Text{max\_Node-}{#1}}}{#2} \Text{ #3}}

\newcommand {\Vgam}[1]{\Index {\gamma}{#1}}
\newcommand {\Egam}[2]{\Equs {\Vgam{#1}}{#2}}

\newcommand {\Emu}[2]{\Equs {{\mu}{#1}}{#2}}
\newcommand {\Esigg}[2]{\Equs {{\sigma}^2{#1}}{#2}}

\newcommand {\Vlambda}[1]{\Index {\lambda}{#1}}

\newcommand {\VV}[1]{\Index {\Vector V}{#1}}
\newcommand {\Vv}[1]{\Index {\Vector v}{#1}}
\newcommand {\Vsv}[1]{\Index {v}{#1}}
\newcommand {\VX}[1]{\Index {\Vector X}{#1}}
\newcommand {\VsX}[1]{\Index {X}{\SI{\Index {}{#1}}}}
\newcommand {\Vx}[1]{\Index {\Vector x}{\SI{\Index {}{#1}}}}
\newcommand {\Vsx}[1]{\Index {x}{\SI{\Index {}{#1}}}}
\newcommand {\VZ}[1]{\Index {\Vector Z}{#1}}
\newcommand {\Vz}[1]{\Index {\Vector z}{\SI{\Index {}{#1}}}}
\newcommand {\Vsz}[1]{\Index {z}{\SI{\Index {}{#1}}}}
\newcommand {\VIndex}[2]{\Index {\Vector {#1}}{#2}}
\newcommand {\VY}[1]{\Index {\Vector Y}{#1}}
\newcommand {\Vy}[1]{\Index {\Vector y}{#1}}
\newcommand {\Vsy}[1]{\Index {y}{\SI{\Index {}{#1}}}}

\newcommand {\VRV}[1]{\Index {\Rdata {\Vector V}}{#1}}
\newcommand {\VRsV}[1]{\Index {\Rdata {V}}{#1}}
\newcommand {\VRX}[1]{\Index {\Rdata {\Vector X}}{#1}}
\newcommand {\VRx}[1]{\Index {\Rdata {\Vector x}}{\SI{\Index {}{#1}}}}
\newcommand {\VRsx}[1]{\Index {\Rdata {x}}{\SI{\Index {}{#1}}}}
\newcommand {\VRZ}[1]{\Index {\Rdata {\Vector Z}}{#1}}
\newcommand {\VRz}[1]{\Index {\Rdata {\Vector z}}{\SI{\Index {}{#1}}}}
\newcommand {\VRsz}[1]{\Index {\Rdata {z}}{\SI{\Index {}{#1}}}}
\newcommand {\VTX}[1]{\Index {\Tdata {\Vector X}}{#1}}
\newcommand {\VTx}[1]{\Index {\Tdata {\Vector x}}{\SI{\Index {}{#1}}}}
\newcommand {\VTsx}[1]{\Index {\Tdata {x}}{\SI{\Index {}{#1}}}}
\newcommand {\VTsX}[1]{\Index {\Tdata {X}}{\SI{\Index {}{#1}}}}
\newcommand {\VTZ}[1]{\Index {\Tdata {\Vector Z}}{#1}}
\newcommand {\VTz}[1]{\Index {\Tdata {\Vector z}}{\SI{\Index {}{#1}}}}
\newcommand {\VTsz}[1]{\Index {\Tdata {z}}{\SI{\Index {}{#1}}}}
\newcommand {\VOG}[1]{\Vector{\Omega}{#1}}

\newcommand {\Sigg}[1]{\Data {{\sigma}^2({#1})}}
\newcommand {\Sig}[1]{\Data {{\sigma}({#1})}}

\newcommand {\Mu}[1]{\Data{{\mu} ({#1})}}
\newcommand {\Eig}[1]{\Data {\lambda}({\Vector {#1}}) }
\newcommand {\Her}[1]{\Power {#1}{\!H}}
\newcommand {\Tra}[1]{\Power {#1}{\!T}}

\newcommand {\VF}[3] {\DataIJ {\Fdata {#1}}{#2}{#3}}
\newcommand {\VRr}[2] {\DataN {\Fdata {#1}}{#2}}

\newcommand {\Tcol}[2] {\multicolumn{1}{#1}{#2} }
\newcommand {\Tcols}[3] {\multicolumn{#1}{#2}{#3} }
\newcommand {\Cur}[2] {\mbox {\Data {#1}-\Data {#2}}}

\newcommand {\VDelta}[1] {\Data {\Delta\!{#1}}}

\newcommand {\STE}[1] {\Fdata {E}{\Data{({#1})}}}
\newcommand {\STD}[1] {\Fdata {D}{\Data{({#1})}}}

\newcommand {\TestF}[1] {\Data {\varphi(#1)}}
\newcommand {\ROMAN}[1] {\uppercase\expandafter{\romannumeral#1}}

\def \FuncC #1#2{
\begin{equation}
{#2}
\label {#1}
\end{equation}
}

\def \FuncCC #1#2#3#4#5#6{
\begin{equation}
#2=
\begin{cases}
    #3 & #4 \\
    #5 & #6
\end{cases}
\label{#1}
\end{equation}
}

\def \Figff #1#2#3#4#5#6#7{   
\begin{figure}[#7]
\centering
\subfloat[#2]{
\label{#1a}
\includegraphics[width=0.23\textwidth]{#4}
}
\subfloat[#3]{
\label{#1b}
\includegraphics[width=0.23\textwidth]{#5}
}
\caption{\small #6}
\label{#1}
\end{figure}
}

\def \Figffb #1#2#3#4#5#6#7#8#9{   
\begin{figure}[#9]
\centering
\subfloat[#2]{
\label{#1a}
\includegraphics[width=0.23\textwidth]{#5}
}
\subfloat[#3]{
\label{#1b}
\includegraphics[width=0.23\textwidth]{#6}
}

\subfloat[{#4}]{
\label{#1c}
\includegraphics[width=0.48\textwidth]{#7}
}
\caption{\small #8}
\label{#1}
\end{figure}
}

\def \Figffp #1#2#3#4#5#6#7{   
\begin{figure*}[#7]
\centering
\subfloat[#2]{
\label{#1a}
\begin{minipage}[t]{0.24\textwidth}
\centering
\includegraphics[width=1\textwidth]{#4}
\end{minipage}
}
\subfloat[#3]{
\label{#1b}
\begin{minipage}[t]{0.24\textwidth}
\centering
\includegraphics[width=1\textwidth]{#5}
\end{minipage}
}
\caption{\small #6}
\label{#1}
\end{figure*}
}

\def \Figf #1#2#3#4{   
\begin{figure}[#4]
\centering
\includegraphics[width=0.48\textwidth]{#2}

\caption{\small #3}
\label{#1}
\end{figure}
}

\definecolor{Orange}{RGB}{249,106,027}
\definecolor{sOrange}{RGB}{251,166,118}
\definecolor{ssOrange}{RGB}{254,213,190}

\definecolor{Blue}{RGB}{008,161,217}
\definecolor{sBlue}{RGB}{090,206,249}
\definecolor{ssBlue}{RGB}{200,239,253}

\title{Short-term Electric Load Forecasting Using TensorFlow and Deep Auto-Encoders}

\author{ Xin Shi$^*$,~\IEEEmembership{Student Member,~IEEE}
\thanks{

$^*$Department of Electrical Engineering, Center for Big Data and Artificial Intelligence, State Energy Smart Grid Research and Development Center, Shanghai Jiaotong University, Shanghai 200240, China.(e-mail: dugushixin@sjtu.edu.cn.)
}
}

\maketitle

\begin{abstract}
This paper conducts research on the short-term
electric load forecast method under the background of big
data. It builds a new electric load forecast model based on
Deep Auto-Encoder Networks (DAENs), which takes into account
multidimensional load-related data sets including historical load
value, temperature, day type, etc. A new distributed short-term
load forecast method based on TensorFlow and DAENs
is therefore proposed, with an algorithm flowchart designed.
This method overcomes the shortcomings of traditional neural
network methods, such as over-fitting, slow convergence and local
optimum, etc. Case study results show that the proposed method
has obvious advantages in prediction accuracy, stability, and
expansibility compared with those based on traditional neural
networks. Thus, this model can better meet the demands of short-term
electric load forecasting under big data scenario.
\end{abstract}


\IEEEpeerreviewmaketitle

\section{Introduction}
\IEEEPARstart{S}{erving} as the basis of the planning and dispatching of power systems, short-term electric load forecasting is the premise of achieving automatic generation control and economic dispatching. Its accuracy has an impact on the safety, stability, and economy of the normal operation of the power systems. An accurate load forecast method can reduce operating costs, keep power markets efficient, and provide a better understanding of the dynamics of the monitored systems\cite{7286732}\cite{5613970}. Therefore, it is of great importance to study short-term load forecast methods.

For a long time, electric load forecast methods have been researched by scholars at home and abroad, which in general fall into two categories, traditional methods and artificial intelligence methods. The former mainly include time series \cite{brockwell2013time}\cite{hagan1987time}, regression \cite{papalexopoulos1990regression} and grey system theory \cite{julong1989introduction}\cite{zhang2004application}, etc. They have simple models and fast calculation speed, but cannot simulate electric load under complicated situations, leading to an unsatisfying prediction accuracy. The latter mainly refer to some machine learning methods, including back propagation neural networks (BPNNs) \cite{hippert2001neural}\cite{tiwari2015comparative}, support vector machine (SVM) \cite{kavousi2014new}\cite{ceperic2013strategy}, extreme learning machine (ELM) \cite{zhang2013short}\cite{chen2012electricity}, etc. The prediction accuracy has been improved to some extent by using those intelligent methods. However, those intelligent methods have limited learning ability and are not capable of transforming and processing large data samples, which can easily lead to over-fitting, slow convergence, local optimum, and many other problems. As a result, the prediction accuracy has certain limitations, which cannot meet the demands and challenges of big data analysis.

With the emergence of big data in the power sector, load-related data dimensions continue to increase, which brings new challenges to the electric load forecast work. We must find a new method to effectively achieve the goal of electric load forecasting and analysis under this scenario. In 2006, Geoffrey Hinton from Canada proposed the method of deep learning\cite{hinton2006reducing}\cite{hinton2006fast}, which opened a new era of deep learning in academic and industrial fields. Deep Auto-Encoder Networks (DAENs), a deep learning method, conducts feature transformation on training samples, layer by layer, to build a machine learning model with multiple hidden layers, projecting the feature representation in the original scenario into a new one, which makes it easier to predict, and ultimately improves prediction accuracy. Compared to methods that construct features by artificial rules, the DAENs method is better at depicting the rich inner information of data, effectively meeting the demands of big data analysis. It has been the hot spot of international machine learning research field\cite{rifai2011contractive}\cite{masci2011stacked}\cite{suk2015latent}, but few attempts have been made to introduce it in the application of electric load forecasting.

\subsection{Contributions}
In this paper, based on multidimensional load-related data such as historical load values, temperature, day types etc., a new distributed short-term load forecast method using TensorFlow and DAENs is proposed. In Sect. \ref{section: background} we give the mathematical background and theoretical foundation. The training processes of DAENs include pre-training and fine-tuning, which are introduced and analyzed in detail. In Sect. \ref{section: model}, we first introduce TensorFlow. Then we construct load forecast model based on DAENs. Considering the complex structure of DAENs and high real-time demand of load forecasting, we designed a parallel algorithm flowchart using the multi-GPU mode analysis framework of TensorFlow, which greatly enhances the calculation speed. Load-related data is processed using the normalization formula, fuzzy membership function, and weighted one-hot encoding method. The evaluation index of load forecasting is defined at the end. In Sect. \ref{section: case}, with load-related data provided by European Smart Technology Network(EUNITE), the forecast method proposed in this paper is validated and its advantages are highlighted. Furthermore, comparison of this method with the method based on BPNNs and ELM is made. In Sect. \ref{section: conclution}, we report our conclusions.

\subsection{Related Works}
Auto-Encoder networks have been widely explored in the past years. Reference \cite{betechuoh2006autoencoder} introduces a new method to analyze the human immunodeficiency virus using a combination of Auto-Encoder networks and genetic algorithms, which outperforms the conventional feedforward neural network models and is a much better classifier. Reference \cite{vincent2008extracting} extracts and composes robust features by using denoising Auto-Encoders, which shows surprising results. In reference \cite{chicco2014deep}, an algorithm that aids the curation of gene annotations and predicts previously-unidentified gene functions is designed by using deep Auto-Encoder neural networks. Experiments show that deep Auto-Encoder neural networks achieve better performance than other standard machine learning methods, including the popular truncated singular value decomposition. Reference \cite{vishnubhotla2010autoencoder} proposes a novel method for modeling the excitation through an Auto-Encoder, which produces speech of higher perceptual quality compared to conventional pulse-excited speech signals. Reference \cite{Msiza2016Autoencoder} uses Auto-Encoder neural networks for water demand predictive modeling. In the research done above, Auto-Encoder networks have been successfully used for solving classification or predication problems.

\section{Mathematical Background and Theoretical Foundation}
\label{section: background}


\subsection{Deep Auto-Encoder Networks}

DAENs are one kind of neural networks composed of multi-layer Auto-Encoders(AE), which are similar to traditional multi-layer neural networks in structure. For example, take one layer of an AE, the network structure is shown in Figure 1.

\begin{figure}[htb]
\centerline{
\includegraphics[width=.20\textheight,height=.35\textwidth]{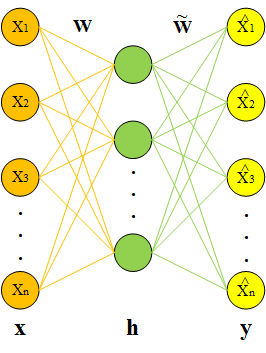}
}
\caption{AE Network Structure.}
\label{fig:AE}
\end{figure}

One basic AE can be viewed as a three-layer traditional neural network, including one input layer, one hidden layer and one output layer. The input layer and output layer are of the same size. Transition from the input layer to the hidden layer is called the encoding process. The transition  from the hidden layer to the output layer is called the decoding process. Assume $f$ and $g$ respectively represent the encoding and decoding function. Then the two processes above can be defined as

\begin{equation}
\label{Eq:h}
\begin{aligned}
  {\mathop{\bf{\emph h}}\nolimits}  = f(x) = {S_f}({\bf{\emph W}}x + {\bf{\emph p}})
\end{aligned},
\end{equation}

\begin{equation}
\label{Eq:y}
\begin{aligned}
{\mathop{\bf{\emph y}}\nolimits}  = g(h) = {S_g}({\bf{\hat{\emph W}}}h + {\bf{\emph q}})
\end{aligned},
\end{equation}

where ${S}_{f}$ and ${S}_{g}$ are the sigmoid functions, ${\bf{\emph W}}$ is the weight matrix between the input layer and the hidden layer, and ${\bf{\hat{\emph W}}}$ is the weight matrix between the hidden layer and the output layer, whose initial value is ${{\bf{\emph W}}^{\mathop{\rm T}\nolimits}}$. ${\bf{\emph p}}$ and ${\bf{\emph q}}$ are the bias vectors of the hidden layers and output layers separately. For simplicity, we use ${\bf{\theta }}$ to represent those parameters, namely ${\bf{\theta  = }}\left\{ {{\bf{\emph W,\emph p,\emph q}}} \right\}$.

Assuming training sample sets as ${\rm{\bf{S}}} = \left\{ {{{\bf{\emph x}}^{(1)}}, {{\bf{\emph x}}^{(2)}},  \cdots , {{\bf{\emph x}}^{(N)}}} \right\}$, we can optimize ${\bf{\theta }}$ through training the AE network on it. The target of training the AE network is to make its output ${\bf{\emph y}}$ and input ${\bf{\emph x}}$ be as equal as possible. The approximation can be described by the reconstruction error function ${{L}({\bf{\emph x,\emph y}})}$, which is defined as

\begin{equation}
\label{Eq:L}
\begin{aligned}
L({\bf{\emph x}},{\bf{\emph y}}) =  - \sum\limits_{i = 1}^n {\left[ {{{\bf{\emph x}}_i}\log ({{\bf{\emph y}}_i}) + (1 - {{\bf{\emph x}}_i})\log (1 - {{\bf{\emph y}}_i})} \right]}
\end{aligned}.
\end{equation}

Combining ${{L}({\bf{\emph x,\emph y}})}$ and ${\bf{S}}$, we can obtain the loss function as
\begin{equation}
\label{Eq:J}
\begin{aligned}
{J_{AE}}({\bf{\theta }}) = \frac{1}{N}\sum\limits_{{\bf{\emph x}} \in {\bf{S}}} {L({\bf{\emph x}},g(f({\bf{\emph x}})))}
\end{aligned}.
\end{equation}

Then we can get ${\bf{\theta }}$ by optimizing ${J_{AE}}\left( {\bf{\theta }} \right)$ through gradient descent methods, such as SGD \cite{SGD}, ADAM \cite{2015Adam}, RMSPROP \cite{RMSprop}, etc. However, it is likely to obtain an identity function in practical application if we optimize the loss function directly. To solve this problem, we can add sparse limits into the loss function. Assuming the input is ${{\rm{\emph x}}^{(i)}}$ and ${h_j}({{\bf{\emph x}}^{(i)}})$ indicates the activation degree of the $j$th neuron in the hidden layer, the average activation  on training sets ${\bf{S}}$ is
\begin{equation}
\label{Eq:rhoj}
\begin{aligned}
{\hat \rho _j} = \frac{1}{N}\sum\limits_{i = 1}^N {{h_j}({{\bf{\emph x}}^{(i)}})}
\end{aligned}.
\end{equation}

In order to guarantee the sparse limits, neurons in the hidden layers are suppressed to zero most of the time, that is
\begin{equation}
\label{Eq:rho}
\begin{aligned}
{\hat \rho _j} = \rho
\end{aligned},
\end{equation}
where ${\rho}$ is the sparse parameter and usually gets a small value. ${\hat \rho _j}$ that is obviously different from ${\rho}$ should be ``punished'' using KL divergence method \cite{yu2013kl}. We can add
$\sum\limits_{j = 1}^m {KL(\rho ||{{\hat \rho }_j})} $ into the loss function. Then the loss function is
\begin{equation}
\label{Eq:rho}
\begin{aligned}
{J_{AE + sp}}({\bf{\theta }}) = \sum\limits_{{\bf{\emph x}} \in {\bf{S}}} {L({\bf{\emph x}},g(f({\bf{\emph x}})))}  + \beta \sum\limits_{j = 1}^m {KL(\rho ||{{\hat \rho }_j})}
\end{aligned},
\end{equation}
where ${\beta }$ is the weight coefficient of sparse penalty term, and
\begin{equation}
\label{Eq:kl}
\begin{aligned}
KL(\rho ||{\hat \rho _j}) = \rho *\log \frac{\rho }{{{{\hat \rho }_j}}} + (1 - \rho )*\log \frac{{1 - \rho }}{{1 - {{\hat \rho }_j}}}
\end{aligned}.
\end{equation}

Equation (8) shows that $\sum\limits_{j = 1}^m {KL(\rho ||{{\hat \rho }_j})} $ gradually decreases as ${\hat \rho _j}$ approaches ${\rho}$, and reaches zero when ${\hat \rho _j}$ equals ${\rho}$. Therefore, we can minimize the difference between ${\hat \rho _j}$ and ${\rho}$ by optimizing the loss function in equation (7).

\subsection{Training}
DAENs is trained layer by layer. The outputs of the previous layer serve as the inputs of next level, which effectively solves the problem of gradient diffusion in traditional neural networks. The entire network training process is divided into two stages: pre-training and fine-tuning .

Pre-training is essentially the process of initializing network parameters layer by layer through unsupervised feature optimization algorithms. In this process, we set the target value equal to the input value, and adopt stochastic gradient descent or similar methods as the training algorithms, which means the output vector of the hidden layer is taken as the input of the next layer for training after one layer of AE training is completed. The steps above are repeated until the training of all AE layers is finished. After pre-training, the labelled data set is used to calibrate the parameters of the whole DAENs to achieve the global optimum by training with stochastic gradient descent or similar methods, which is called fine-tuning. The method combining pre-training with fine-tuning usually outperforms those based on traditional neural networks, because it can effectively avoid slow convergence, over-fitting, local optimum, and instability problems.

\section{Load Forecasting Based on DAENs and TensorFlow}
\label{section: model}

\subsection{Introduction to TensorFlow}
 TensorFlow is an open source artificial intelligence programming system released by Google in November, 2015 \cite{abadi2015tensorflow}. It can transmit data of complex structure to artificial neural networks for process and analysis. It has been widely used in deep machine learning areas, such as speech recognition \cite{abadi2016tensorflow}\cite{sharma2016asr}, image processing \cite{kurakin2016adversarial}\cite{rampasek2016tensorflow}, natural language processing \cite{goth2016deep}\cite{hewlett2016wikireading}, etc. It runs on devices ranging from a small smartphone to thousands of data center servers. TensorFlow conducts calculation in the form of data flow graphs and this flexible architecture allows it to support single or multiple CPU or GPU computing, which greatly improves the ability of data computation and analysis in big data case. Based on TensorFlow, this paper combines the DAENs model and TensorFlow distributed data analysis framework to realize the distributed short-term load forecasting of power system.

 TensorFlow provides the following key concepts for data flow graph calculation.
 \begin{itemize}
\item Graph: It is used to represent the calculation task and describe the calculation process.
\item Op: The acronym for operation, which represents the nodes in the graph and is responsible for performing the computation.
\item Session: It is responsible for the implementation of the graph, namely distributing the ops to CPU or GPU devices and providing op implementation methods.
\item Tensor: It is used to indicate data.
\item Variable: It is used to maintain the states.
\item Feed/Fetch: It is used to input or output data for any operation.
\end{itemize}

\subsection{Load Forecast Model Based on DAENs and TensorFlow}
The load forecast model based on DAENs constructed in this paper is shown in Fig.2. The DAENs consist of all layers except for the last one. The number of layers is determined by the dimension and volume of input data. High dimension and large volume of data will lead to the increasing of AE layers, and vice versa. We add an output layer at the end of the model to serve as the forecast layer, where a linear regression model is used. The input of the model is multidimensional load-related data including historical load value, temperature, day types, etc., and the output is the load value at certain time on the forecast day. What is important about the model is that it is scalable, which means we can achieve good performance of the model through adding the number of network layers and neurons when dimension and volume of sample data increase, because its training method overcomes gradient diffusion while traditional machine learning methods do not.
\begin{figure}[htb]
\centerline{
\includegraphics[width=.35\textheight,height=.22\textwidth]{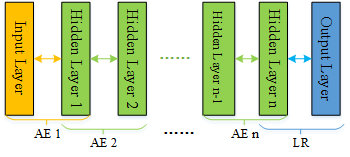}
}
\caption{Load Forecasting Model Based on DAENs.}
\label{fig:DAENs}
\end{figure}

 A parallel algorithm flow chart is designed, shown in Fig. 3, that combines the DAENs-based load forecast model with multi-GPU mode analysis framework of TensorFlow.
\begin{figure*}[htb]
\centerline{
\includegraphics[width=.75\textheight,height=.45\textwidth]{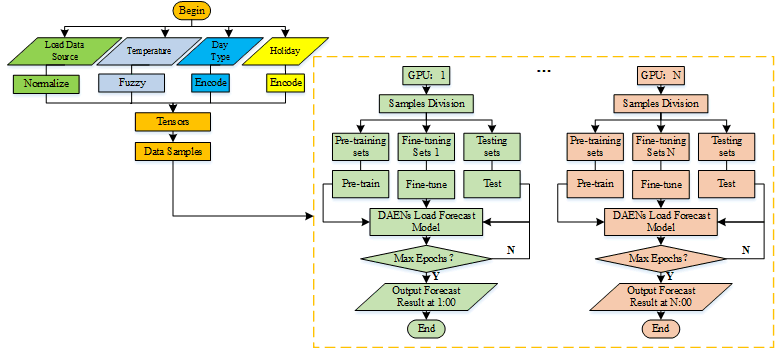}
}
\caption{Parallel Algorithm Flow Chart Based on DAENs and TensorFlow.}
\label{fig:multigpu}
\end{figure*}

As displayed in Fig. 3, the load-related data is processed and loaded into TensorFlow's tensors to form data samples. Then data samples are distributed into GPUs by the computer CPU and further divided into pre-training sets, fine-tuning sets, and testing sets for pre-training, fine-tuning, and testing the forecast model respectively. Load forecast value at different moments of the forecast day are accomplished by parallel model training of multiple GPUs. The load-related data is defined and processed as follows.

1) Electric load forecasting is related to multiple factors, including historical load value, daily temperature, day types, etc. Here we define the load-related dataset as ${\rm{\bf{S}}} = \left\{ {\bf{L}},{\bf{T}},{\bf{D}},{\bf{H}} \right\}$. {\bf{L}} is the source load data set,{\bf{T}} is the daily average temperature data set, {\bf{D}} is the day type data set, and {\bf{H}} is the holiday data set.

2) Source load data are normalized according to equation (9).
\begin{equation}
\label{Eq:L}
\begin{aligned}
L = \frac{{{L_t} - {L_{\min }}}}{{{L_{\max }} - {L_{\min }}}}
\end{aligned},
\end{equation}
where ${L_{\max}}$ and ${L_{\min}}$ respectively represent the maximum and minimum load value in the training sample sets, ${L_t}$ represents the current load value, and $L$ represents the normalized load value.

3) Fuzzy membership functions of low, medium, and high temperature are defined to blur temperature data to be 0, 0.25, and 0.5 separately, which are shown as equation (10), (11) and (12).
\begin{equation}
\label{Eq:P1}
\begin{aligned}
{p_1} = \left\{ \begin{array}{l}
0\;\;\;\;\;\;\;\;\;t \ge 10\\
\frac{{10{\rm{ - }}t}}{{15}}\;\;\; - 5 \le t \le 10\\
1\;\;\;\;\;\;\;\;\;t \le  - 5
\end{array} \right.
\end{aligned},
\end{equation}

\begin{equation}
\label{Eq:P2}
\begin{aligned}
{p_2} = \left\{ \begin{array}{l}
0\;\;\;\;\;\;\;\;\;t \le 0,t \ge 20\\
\frac{t}{{10}}\;\;\;\;\;\;\;0 \le t \le 10\\
\frac{{20{\rm{ - }}t}}{{10}}\;\;\;\;10 \le \;t \le 20
\end{array} \right.
\end{aligned},
\end{equation}

\begin{equation}
\label{Eq:P3}
\begin{aligned}
{p_3} = \left\{ \begin{array}{l}
0\;\;\;\;\;\;\;\;\;t \le 15\\
\frac{{t{\rm{ - }}10}}{{20}}\;\;\;\;15 \le t \le 30\\
1\;\;\;\;\;\;\;\;\;t \ge 30
\end{array} \right.
\end{aligned}.
\end{equation}

In the equations above, ${t}$ is the daily average temperature value and ${P_{i}}$ is the probability value.

4) Day type here is used to distinguish weekday attributes by using a weighted one-hot encoding method \cite{murphy2012machine}. The weight value is 0.5. The encoded data is shown in Table I.
\begin{table}[htbp]
\caption{Day Type Encoding.}
\label{Tab: Case0LESs}
\centering

\begin{minipage}[!h]{0.48\textwidth}
\centering

\footnotesize
\begin{tabular}{p{2.4cm}p{2.4cm}}   
\toprule[1.5pt]
\textbf {Date} & \textbf{Encoded Value}\\
\toprule[1pt]
Monday & $(0,0,0,0,0,0,0.5)$ \\
\hline

Tuesday & $(0,0,0,0,0,0.5,0)$ \\
\hline

Wednesday & $(0,0,0,0,0.5,0,0)$ \\
\hline

Thursday & $(0,0,0,0.5,0,0,0)$ \\
\hline

Friday & $(0,0,0.5,0,0,0,0)$ \\
\hline

Saturday & $(0,0.5,0,0,0,0,0)$ \\
\hline

Sunday & $(0.5,0,0,0,0,0,0)$ \\
\hline
\toprule[1pt]
\end{tabular}
\end{minipage}
\end{table}

5) Holiday falls into either the ``yes'' or ``no'' category here, which is encoded as 0.5 or 0 respectively.

\subsection{Evaluation Index }
Prediction error is an important index to evaluate the performance of one forecast method. In this paper, maximum relative error ($MaxRe$), minimum relative error ($MinRe$), and mean absolute error ($Mae$) are used to evaluate the prediction performance of the method proposed, which are separately defined as
\begin{equation}
\label{Eq:maxre}
\begin{aligned}
MaxRe = \max (\left| {\frac{{{F_i} - {{\hat F}_i}}}{{{F_i}}}} \right|.100\% )
\end{aligned},
\end{equation}

\begin{equation}
\label{Eq:maxre}
\begin{aligned}
MinRe = \min (\left| {\frac{{{F_i} - {{\hat F}_i}}}{{{F_i}}}} \right|.100\% )
\end{aligned},
\end{equation}

\begin{equation}
\label{Eq:maxre}
\begin{aligned}
Mae = \frac{1}{n}\sum\limits_{i = 1}^n {(\left| {\frac{{{F_i} - {{\hat F}_i}}}{{{F_i}}}} \right|.100\% )}
\end{aligned},
\end{equation}

where ${F_i}$ is the actual load value and ${\hat F_i}$ is the forecast load value at time $i$. We can conclude from the formulas above that, $MaxRe$ and $MinRe$ can reflect the stability and reliability of the prediction model, while $Mae$ reflects the prediction accuracy of it.

\section{Case Studies}
\label{section: case}

\subsection{Data Samples of the Case}
Based on the load-related data provided by EUNITE, the short-term electric load forecast method proposed in this paper is employed to predict the load data at moments 1:00, 2:00,..., 24:00 on December 31st. We set up 24 load forecast models in total. There are 57 items for each of the model inputs: 1 to 48 for electric loads at moments 0:30, 1:00, 1:30,...,24:00 on a historical day; 49 for the fuzzy value of the historical daily average temperature; 50 to 56 for the day type encoded value of the historical day; 57 for the encoded value of the holiday. The output of each model is the forecast load value at one certain moment, such as 1:00, 2:00, 3:00, etc. The data set is divided as follows for the training process: sample data from January 1st to November 30th is used for pre-training while from November 1st to 23th is used for fine-tuning, with a total of 20169 items. For the testing set, we choose sample data from December 24th to 31st. The data sample format is shown in Table II.

\begin{table*}[htbp]
\caption{Data Samples.}
\label{Tab: Case0LESs}
\centering

\begin{minipage}[!h]{0.98\textwidth}
\centering

\footnotesize
\begin{tabular}{p{1.8cm}p{1.2cm}p{1.2cm}p{1.0cm}p{1.0cm}
p{1.0cm}p{1.0cm}p{1.0cm}p{1.2cm}p{1.2cm}p{1.2cm}}   
\toprule[1.5pt]
  \text{Samples} &   \text{Month} & \text{Day} & {00:30} & {01:00}  & {...} & {23:30} & {24:00} & \text{Temperature} & \text{Day Type} & \text{Holiday}\\
\toprule[1pt]
\multicolumn{7}{l} {$\textbf{Pretraining Set}$}\\
\hline
&1&1&797&794&...&692&686&-7.6&Wednesday&yes\\
&.&.&.&.&...&.&.&.&.&.\\
X=Y&.&.&.&.&...&.&.&.&.&.\\
&.&.&.&.&...&.&.&.&.&.\\
&11&30&665&640&...&617&623&7.2&Sunday&no\\

\toprule[1pt]

\multicolumn{7}{l} {$\textbf{Fine-tuning Set}$ }\\
\hline
&12&1&637&626&...&613&644&6.7&Monday&no\\
&.&.&.&.&...&.&.&.&.&.\\
X&.&.&.&.&...&.&.&.&.&.\\
&.&.&.&.&...&.&.&.&.&.\\
&12&23&674&664&...&640&646&2.4&Tuesday&no\\
\hline
&12&8&---&677&...&---&710&---&---&---\\
&.&.&.&.&...&.&.&.&.&.\\
Y&.&.&.&.&...&.&.&.&.&.\\
&.&.&.&.&...&.&.&.&.&.\\
&12&30&---&673&...&---&663&---&---&---\\

\toprule[1pt]

\multicolumn{7}{l} {$\textbf{Testing Set}$ }\\
\hline
&12&24&650&652&...&621&645&1.1&Wednesday&yes\\
&.&.&.&.&...&.&.&.&.&.\\
X&.&.&.&.&...&.&.&.&.&.\\
&.&.&.&.&...&.&.&.&.&.\\
&12&30&678&673&...&663&663&-0.2&Tuesday&no\\
\hline
Y&12&31&---&668&...&---&692&---&---&---\\

\toprule[1pt]
\end{tabular}
\end{minipage}
\end{table*}

\subsection{Experimental Design and Implementation}
The experiment was conducted on a DELL T430 server, which provides multiple PCIE3.0x16 interfaces for expansion, effectively supporting parallel computing of multi-GPU mode. We mainly used ``Ubuntu14.04'' and ``TensorFlow-0.9.0'' softwares. The experimental parameters are set as follows.
\begin{itemize}
\item AE layers: 2.
\item Number of neurons for each layer: 57, 24, 12, 1.
\item Initial learning rate: 0.01.
\item Maximum pre-training iteration: 2000.
\item Maximum fine-tuning iteration: 250.
\end{itemize}

Tensorboard, a visualization tool of TensorFlow, can display the data flow graph of the model, which is shown in Fig. 4.
\begin{figure}[htb]
\includegraphics[width=.35\textheight,height=.42\textwidth]{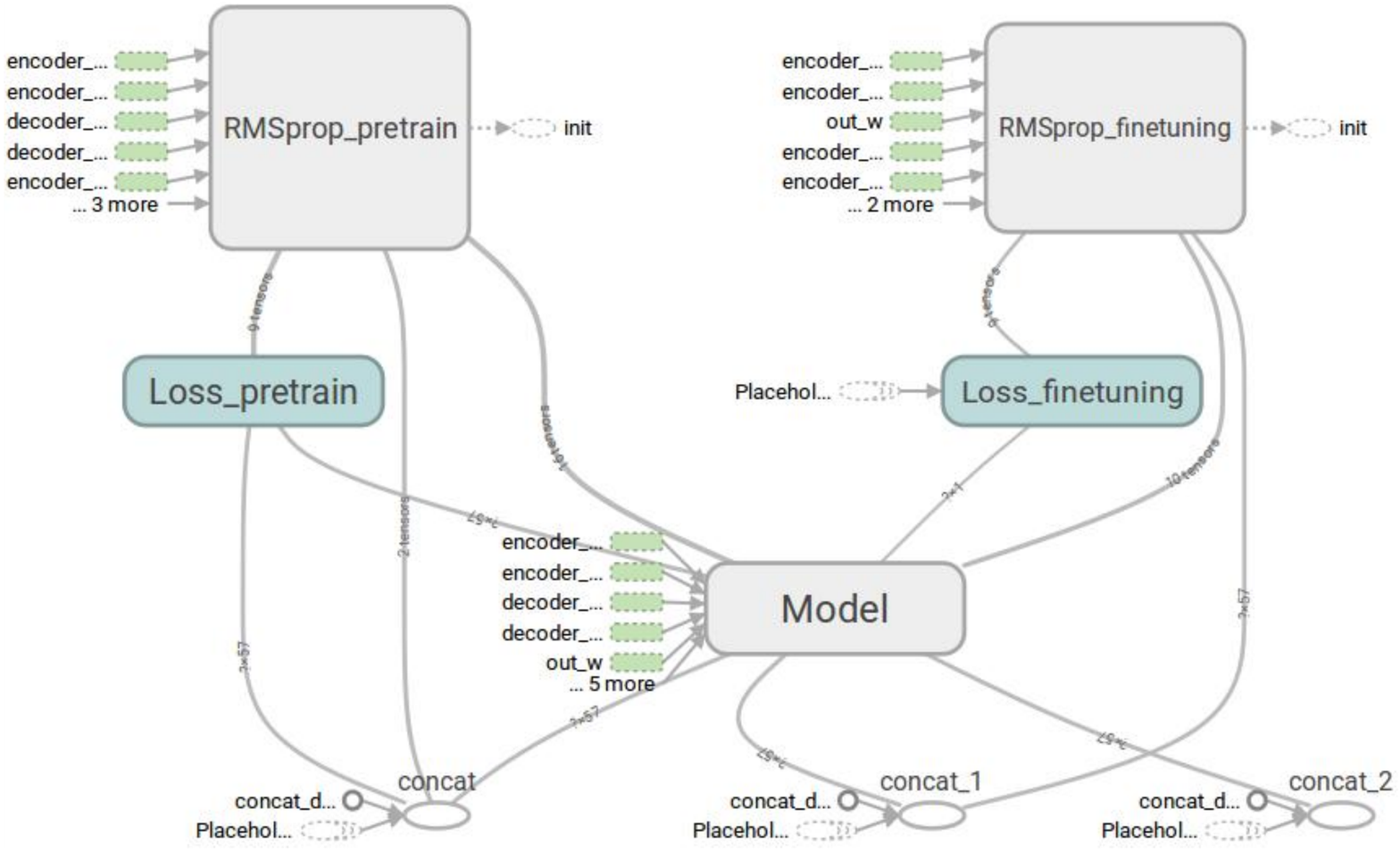}
\caption{Network Data Flow Graph.}
\label{fig:model}
\end{figure}

In Fig. 4, the solid line represents data dependency, while the dashed line represents control dependency. ``${\mathop{\rm Loss}\nolimits} \_{\rm pretrain} $'' and ``${\mathop{\rm RMSprop}\nolimits} \_{\rm pretrain}$'' modules represent the network pre-training process, ``${\mathop{\rm Loss}\nolimits} \_{\rm finetuning}$'' and ``${\mathop{\rm RMSprop}\nolimits} \_{\rm finetuning}$'' modules represent the network fine-tuning process, and ``$\rm Model$'' module represents the network structure. Each module in the figure can be expanded by clicking on the ``$\rm +$'' button in the upper right corner. The expansion form of  ``$\rm Model$'', for example, shows the data flow and operation process, which is displayed in Fig. 5.
\begin{figure}[htb]
\includegraphics[width=.35\textheight,height=.38\textwidth]{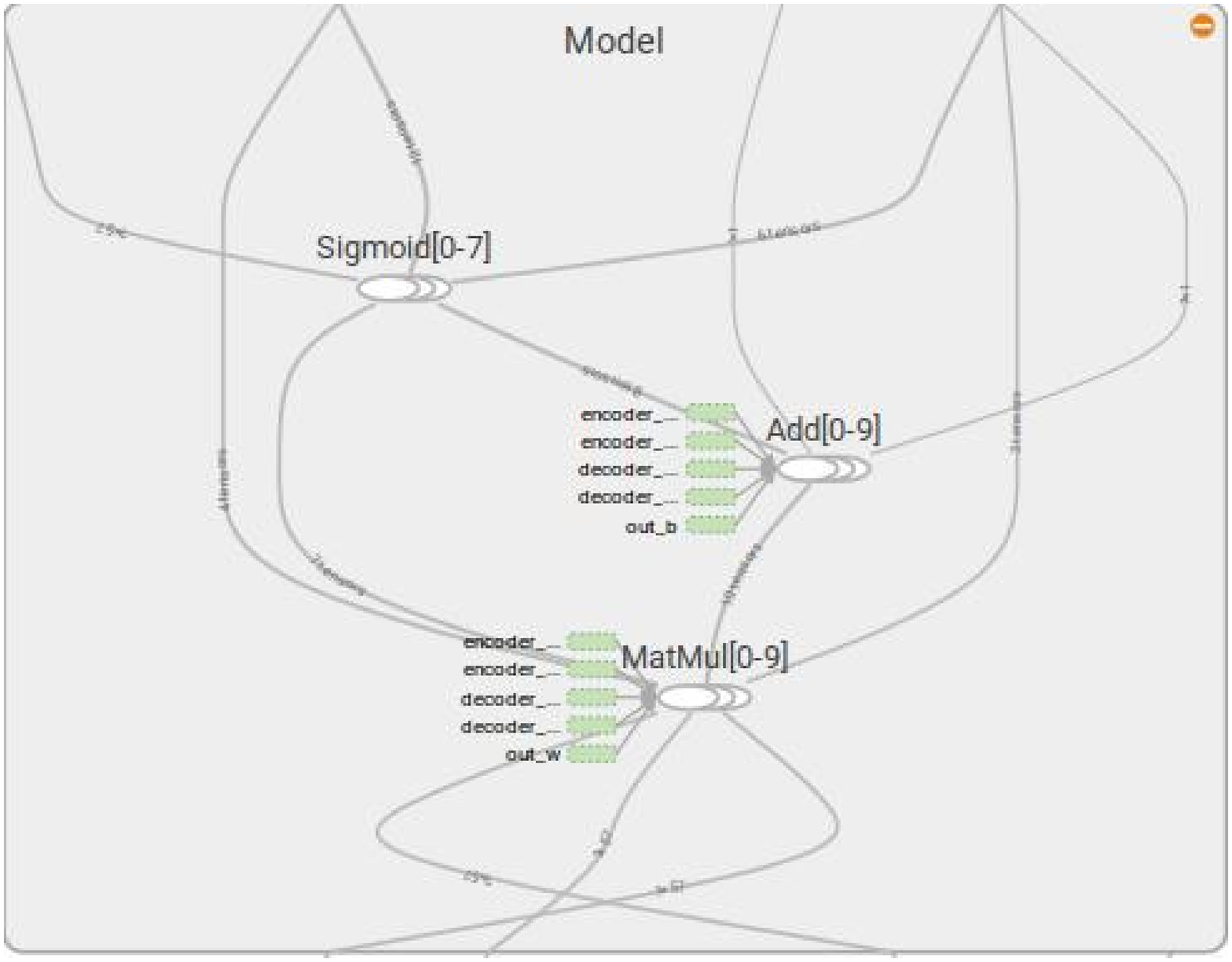}
\caption{Expansion of ``$\rm Model$'' Module.}
\label{fig:expansion}
\end{figure}

Function $tf.scalar\_summary()$ in TensorFlow can be used to trace the error function of pre-training and fine-tuning. Taking GPU 1 for example, the tracing results are separately shown in Fig. 6 and Fig. 7. As can be seen from the figures, errors of pre-training and fine-tuning are continuously decreasing with the increase of iteration times, nearing 0 in the end. The performance of the pre-trained network quickly converges to global optimum with only a few training iterations, which shows that the proposed forecast method has good convergence.
\begin{figure}[htb]
\includegraphics[width=.35\textheight,height=.35\textwidth]{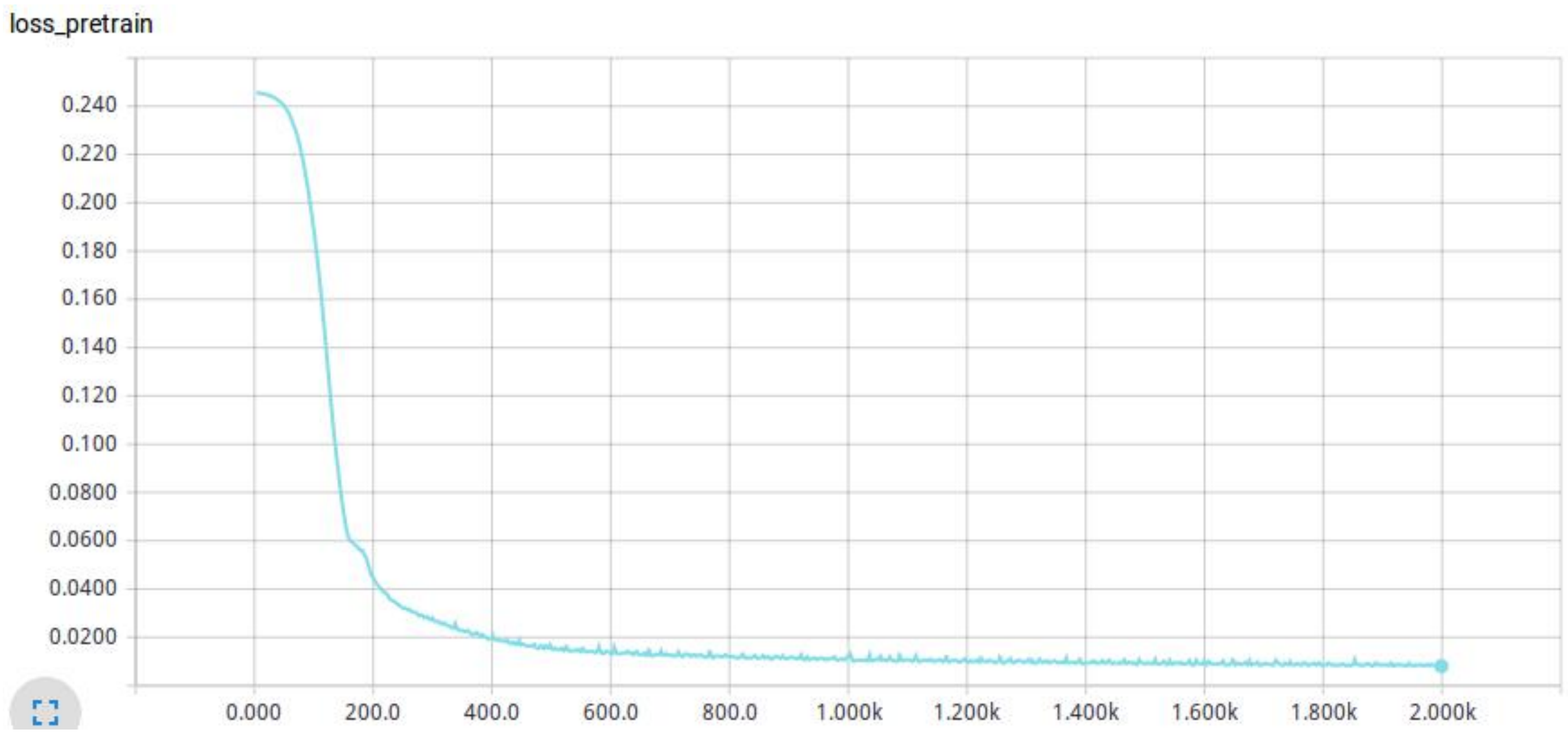}
\caption{Error Variation in Pre-training. The horizontal coordinate denotes the number of training iterations in pre-training and the vertical coordinate denotes the reconstruction error.}
\label{fig:loss_pretrain}
\end{figure}
\begin{figure}[htb]
\includegraphics[width=.35\textheight,height=.35\textwidth]{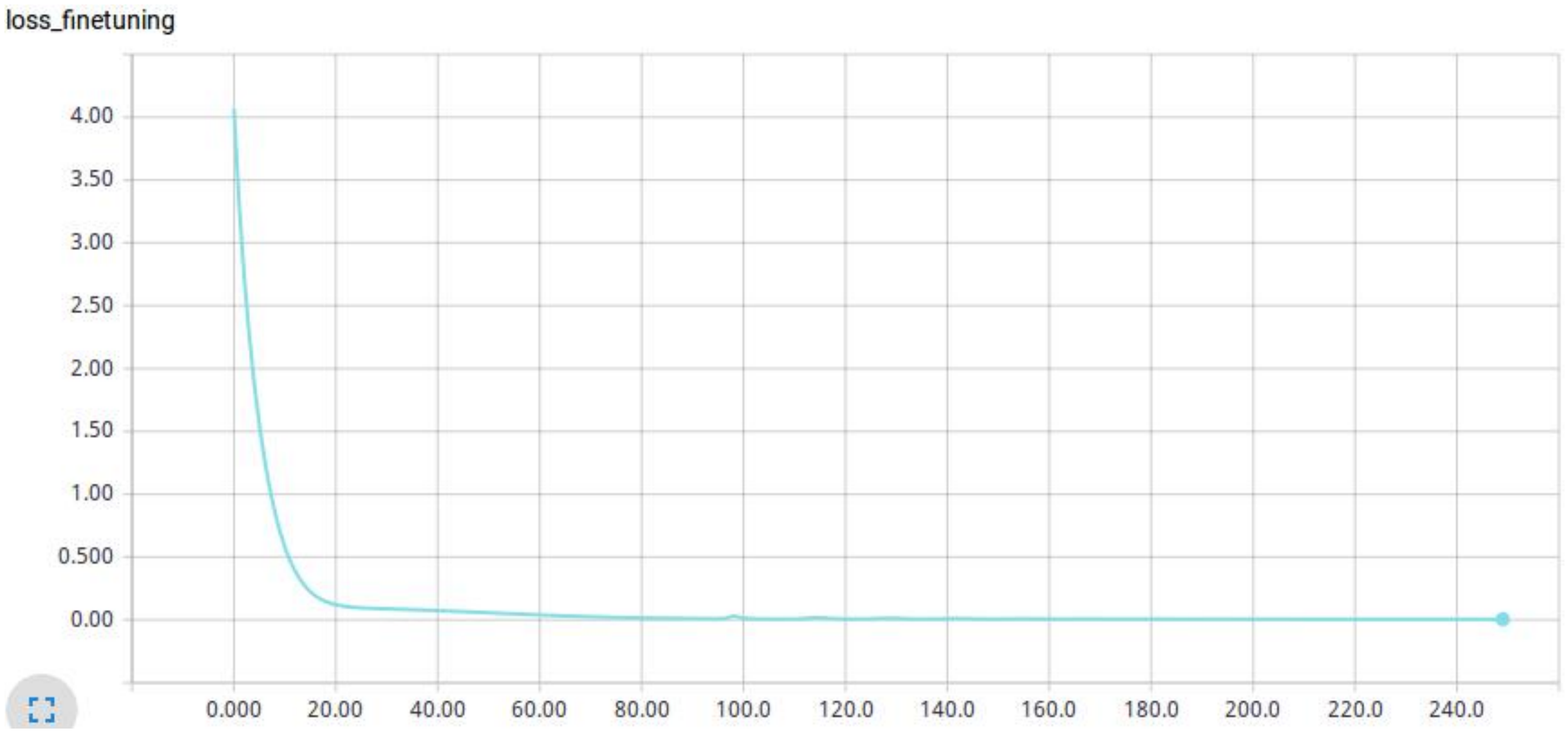}
\caption{Error Variation in Fine-tuning. The horizontal coordinate denotes the number of training iterations in fine-tuning and the vertical coordinate denotes the error.}
\label{fig:loss_finetuning}
\end{figure}

\subsection{Comparison and Analytics of Experimental Results}
In the experiment, we conduct load forecasting on the forecast day by using the load-related data of its previous $1$ to $n$ day. Assuming the forecast results are ${F_1}$, ${F_2}$, ... , ${F_n}$, we calculate the final load forecast value $F$ according to equation (16).
\begin{equation}
\label{Eq:maxre}
\begin{aligned}
F = \sum\limits_{k = 1}^n {{e^{ - \alpha k}}{F_k}}
\end{aligned},
\end{equation}
where $n$ is 7 and $\alpha $ is 0.69 through empirical data in the fine-tuning process. The equation indicates that for each historical day, the closer it is to the forecast day, the more it will contribute to the forecast result.

Comparison of the load forecast value (FV) and the actual value (AV) is shown in Table III, and we calculate the relative error (Re) at each moment, which is also included in Table III.
\begin{table}[htbp]
\caption{Comparison of Load Forecast Value and Actual Value.}
\label{Tab: Comparison}
\centering

\begin{minipage}[!h]{0.48\textwidth}
\centering

\footnotesize
\begin{tabular}{p{1.5cm}p{1.5cm}p{1.5cm}p{1.5cm}}   
\toprule[1.5pt]
  \text{Moments} &   \text{AV/MW} & \text{FV/MW} &  \text{Re/${\rm{\% }}$} \\
\toprule[1pt]
1:00&668&673.3406&0.7995\\
\hline
2:00&642&650.3174&1.2955\\
\hline
3:00&623&625.4593&0.3948\\
\hline
4:00&602&620.9128&3.1417\\
\hline
5:00&618&621.4146&0.5525\\
\hline
6:00&619&611.1831&1.2628\\
\hline
7:00&599&601.7022&0.4511\\
\hline
8:00&602&604.8622&0.4754\\
\hline
9:00&643&640.9628&0.3168\\
\hline
10:00&669&655.8308&1.9685\\
\hline
11:00&700&679.4819&2.9312\\
\hline
12:00&688&689.8511&0.2691\\
\hline
13:00&710&700.154&1.3868\\
\hline
14:00&717&713.8212&0.4433\\
\hline
15:00&700&706.6741&0.9534\\
\hline
16:00&681&687.7114&0.9855\\
\hline
17:00&704&691.7314&1.7427\\
\hline
18:00&691&684.2487&0.9770\\
\hline
19:00&692&688.7641&0.4676\\
\hline
20:00&700&705.1727&0.7390\\
\hline
21:00&648&675.6137&4.2614\\
\hline
22:00&635&656.4427&3.3768\\
\hline
23:00&681&689.8683&1.3022\\
\hline
24:00&692&690.3376&0.2402\\
\toprule[1.5pt]
\end{tabular}
\end{minipage}
\end{table}

As can be seen from Table III, the relative error can be maintained at about $1{\rm{\% }}$ at most moments, the maximum is $4.26{\rm{\% }}$, and the minimum is $0.24{\rm{\% }}$. Besides, the forecast error fluctuation is relatively small. All of these indicate that the overall forecast effect is ideal.

In contrast, we have conducted experiments using load forecasting method based on BPNNs and ELM. 24 models are set up for each method to forecast the load value at 1:00, 2:00,  . . . , 24:00. Because pre-training is not required for them, we just choose the fine-tuning set and testing set separately as training set and testing set in the experiment. Experimental parameters of BPNNs are set as follows.
\begin{itemize}
\item Number of neurons for each layer: 57, 24, 1.
\item Initial learning rate: 0.01.
\item Maximum pre-training iteration: 2000.
\end{itemize}
For ELM, number of neurons for network input layer, hidden layer, and output layer are separately set as 57, 30, and 1.

Comparison of the forecast value and the actual one is shown in Fig. 8.
\begin{figure}[htb]
\includegraphics[width=.35\textheight,height=.35\textwidth]{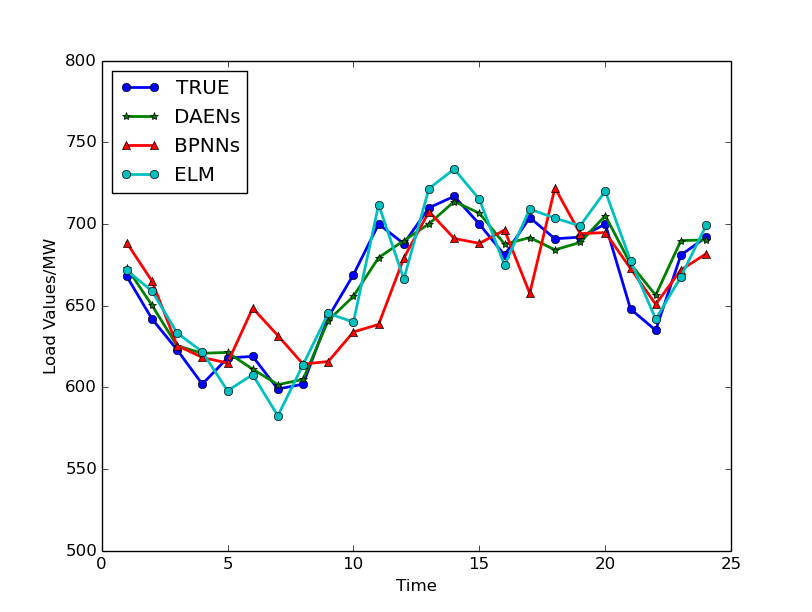}
\caption{Comparison of the Forecast Value and the Actual One.}
\label{fig:comparison}
\end{figure}
It can be seen from Fig. 8 that compared with the actual load value, the load forecast result of the DAENs-based method is more accurate and stable than that of the BPNNs-based and ELM-based method. The comparison also shows that the DAENs-based method can effectively avoid the situation where the BPNNs-based method can easily fall into local optimum. As shown at 17:00 and 18:00 in the figure, compared with the actual load value, there occurs a large deviation in BPNNs-based prediction result, which is likely to result from falling into the local optimal results.

Cumulative distribution function (CDF) of the forecast relative error based on DAENs, BPNNs, and ELM is shown in Fig. 9.
\begin{figure}[htb]
\includegraphics[width=.35\textheight,height=.35\textwidth]{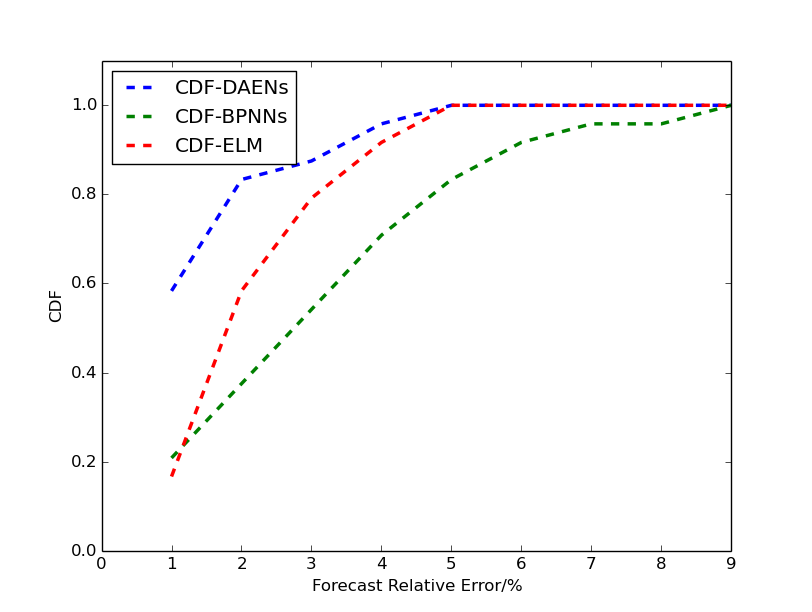}
\caption{The CDF of the Forecast Relative Error Based on DAENs, BPNNs, and ELM.}
\label{fig:cdf}
\end{figure}
We can see that the CDF of the DAENs-based forecast relative error lower than 1\%, 2\% and 3\% is about 0.6, 0.85 and 0.9 respectively. The CDF of the BPNNs-based forecast relative error lower than 1\%, 2\% and 3\% is separately only about 0.2, 0.3 and 0.4, and that of the ELM-based forecast relative error lower than 1\%, 2\% and 3\% is separately about 0.15, 0.6 and 0.75. Obviously, the CDF of the DAENs-based forecast relative error is the highest among them in any error range, which shows that the DAENs-based forecast method is more reliable than the other two methods. Further comparison of the DAENs-based method , the BPNNs-based method, and the ELM-based method is made in terms of $MaxRe$, $MinRe$ and $Mae$. The results are shown in Table IV.
\begin{table}[htbp]
\caption{The Indicator Comparison of Different Load Forecast Methods.}
\label{Tab: Case0LESs}
\centering

\begin{minipage}[!h]{0.48\textwidth}
\centering

\footnotesize
\begin{tabular}{p{1.7cm}p{1.7cm}p{1.7cm}p{1.7cm}}   
\toprule[1.5pt]
  \text{Indicators} & \text{DAENs/${\rm{\% }}$}&   \text{BPNNs/${\rm{\% }}$}& \text{ELM/${\rm{\% }}$}  \\
\toprule[1pt]
$MaxRe$&4.26&8.78&4.57\\
$MinRe$& 0.24&0.31 & 0.36\\
$Mae$& 1.28&2.97 & 2.06\\
\toprule[1.5pt]
\end{tabular}
\end{minipage}
\end{table}

It can be seen from Table IV that the $MaxRe$ of the DAENs-based method is almost 2 times lower than that of the BPNNs-based method, and the $MinRe$ of the former is also lower than that of the the BPNNs-based and the ELM-based method. The $Mae$ of the DAENs-based method is $1.28{\rm{\% }}$, which is nearly 2 times lower than that of the other two methods. It shows that the DAENs-based load forecast method outperforms the BPNNs-based method and the ELM-based method in predicting accuracy and reliability.

\section{Conclusion}
\label{section: conclution}
\normalsize{}
With the emergence of big data in the power sector, traditional load forecast methods present some limitations, such as over-fitting, slow convergence, local optimum, etc., resulting in inadequate forecast accuracy. To enhance load forecast accuracy in a big data scenario, a new distributed short-term load forecast method based on TensorFlow and DAENs by taking advantage of multidimensional load-related data is proposed, and its performance is tested and analyzed. The work of this paper is summarized as follows.

1) The pre-training and fine-tuning process of the DAENs-based load forecast method are separately unsupervised learning and supervised learning. A combination of the two processes makes it possible to overcome the shortcomings of traditional methods, such as slow convergence, over-fitting, local optimum, and instability, which contributes to a more accurate, stable, and reliable forecast result.

2) The DAENs-based load forecast method can easily tackle high-dimension and large-scale data by expanding its network structure, e.g. increasing the number of network layers and neurons. It has good expansibility and can thus meet the requirements for load forecast accuracy under the background of big data.

3) When applied in a big data scenario, the network structure of the DAENs-based model is complex, usually involving a huge amount of computation. By combining the TensorFlow multi-GPU framework, we realize distributed parallel computing to speed up the process of data calculation and analysis and improve work efficiency.

\small{}
\bibliographystyle{IEEEtran}
\bibliography{helx}

\normalsize{}
\end{document}